\documentclass[12pt, preprint]{aastex}

\slugcomment{published in \emph{Journal of Geophysical Research -- Planets}, \textbf{114}: E04007 (2009) | doi: 10.1029/2008JE003254}

\shorttitle{Saturn's Atmosphere Observed by VIMS}
\shortauthors{Choi et al.}

\begin{document}

\title{Cloud Features and Zonal Wind Measurements of Saturn's Atmosphere as Observed by Cassini/VIMS}

\author{David S. Choi, Adam P. Showman, and Robert H. Brown}
\affil{Department of Planetary Sciences, University of Arizona, Tucson, AZ 85721}
\email{dchoi@lpl.arizona.edu}

\begin{abstract}
We present an analysis of data about Saturn's atmosphere from \emph{Cassini}'s Visual and Infrared Mapping Spectrometer (VIMS), focusing on the meteorology of the features seen in the 5-micron spectral window. We present VIMS mosaics and discuss the morphology and general characteristics of the features backlit by Saturn's thermal emission. We have also constructed a zonal wind profile from VIMS feature tracking observation sequences using an automated cloud feature tracker. Comparison with previously constructed profiles from \emph{Voyager} and \emph{Cassini} imaging data reveals broad similarities, suggesting minimal vertical shear of the zonal wind. However, areas of apparent wind shear are present in the VIMS zonal wind profile at jet stream cores. In particular, our analysis shows that the equatorial jet reaches speeds exceeding 450 m s$^{-1}$, similar to speeds obtained during the \emph{Voyager} era. This suggests that recent inferences of relatively slower jet speeds of $\sim$275-375 m s$^{-1}$ are confined to the upper troposphere and that the deep ($>$1 bar) jet has not experienced a significant slowdown. Our measurements of the numerous dark, spotted features seen in the VIMS mosaics reveals that most of these features have diameters less than 1000 km and reside in confined zonal bands between jet stream cores.  We propose that these spot features are vortices and that VIMS and ISS are sensing the same vortices at two different pressure levels. The local structure at the zonal jet streams remains complex, as VIMS may be sensing cloud features that are deeper than the NH$_3$ cloud deck.
\end{abstract}

\keywords{Saturn, Atmospheres --- Atmospheres, Dynamics --- Atmospheres, Structure}

\section{Introduction}

Saturn's atmosphere has a reputation for being bland and unremarkable, especially when compared against Jupiter's colorful bands and vortices. A strongly scattering haze confined to the higher altitudes of the atmosphere is mostly responsible for obscuring the activity occurring at depth. Recent imaging results from \emph{Cassini} have revealed this activity: dozens of compact, dark spots (likely anticylconic vortices) populate Saturn's mid-latitudes \citep{Porco05}, and vigorous storms with inferred lightning discharges imply energetic updrafts in tandem with strong precipitation \citep{Dyudina07}. \citet{Vasavada06} describe tilted cloud streaks, abundant vortices, patchy bright clouds, and dark bands coincident with the location of zonal jet streams throughout Saturn's southern hemisphere. Their observed vortices had lifetimes on the timescale of several to hundreds of days, demonstrating the activity of the atmosphere. Though \emph{Cassini}'s Imaging Science Subsystem (ISS) was not yet able to fully observe the northern hemisphere, \citet{Vasavada06} noted similarities between both hemispheres based on previous studies from \emph{Voyager} [\citet{Smith81}; \citet{Smith82}; \citet{Sromovsky83}].

Saturn's zonal wind profile, especially at its equatorial region, has been the subject of much discussion. \emph{Voyager} imaging data indicated that the broad equatorial jet flowed as fast as 470 m s$^{-1}$ at its peak \citep{Sanchez-Lavega00}. However, Hubble Space Telescope (HST) observations of the jet from 1996--2002 showed evidence for a dramatic slowdown in the velocity of the jet by almost a factor of two \citep{Sanchez-Lavega03} over a broad ($\sim$30$^{\circ}$) span of latitudes. Explanations for the slowdown include a true decrease in the speed of the jet or that HST observed higher altitude clouds that would flow at slower speeds \citep{Flasar05}. \citet{Sayanagi07} performed numerical experiments that suggested that a combination of both is necessary to account for the apparent slowdown. Recent results from the \emph{Cassini} orbiter made by \citet{Porco05} have bridged the \emph{Voyager} and HST measurements, as they supply a measurement of winds near 375 m s$^{-1}$ at the equator using a near-infrared continuum band that samples deeper in the atmosphere than the HST measurements. 

With the arrival of \emph{Cassini} in 2004, planetary scientists gained the capability of using an imaging spectrometer to assess the characteristics of the Saturnian atmosphere. The Visual and Infrared Mapping Spectrometer (VIMS) instrument is a 64x64 pixel spectrograph capable of near-simultaneous observations of a target from 0.3--5 $\mu$m. VIMS is a hybrid instrument combining a slit-scanning visible spectrometer \citep{Capaccioni98} with a spot-scanning near-infrared spectrometer \citep{Brown04}. For this paper, we focus primarily on data in the near-infrared, specifically at 5 $\mu$m. We utilize this spectral window because of its potential to probe deeper layers of the atmosphere. In 5 $\mu$m, Saturn's thermal emission can be directly observed, and Saturn's atmospheric cloud features are seen ``backlit" against the infrared emission, indicating that these features are located anywhere in between the top of the atmosphere and the 5 $\mu$m emission level. \citet{Bjoraker07} estimates the 5 $\mu$m emission layer at around 5 bars, which is below the estimated altitude (1-2 bars) for the cloud tops seen in ISS images observing reflected solar radiation. 

\citet{Baines05EMP} and \citet{Baines06} conducted an initial study of results from the first year of data from the \emph{Cassini} VIMS instrument. Although their initial reports focused mostly on constraining atmospheric compositions, the reports made the important initial discovery of meteorological features seen in Saturn's atmosphere in the 5 $\mu$m spectral region. \citet{Momary06} characterize these features, some of which have peculiar descriptions: certain features in the northern high latitudes are ``donut-like'' and are presumably vortices with a relatively cloud-less inner core. Furthermore, a train of cloud clearings arranged like a ``string of pearls'' inhabits the northern mid-latitudes. They determine that these clearings are quasi-evenly spaced with a near 4:1 flux contrast ratio and estimate their altitude to be near the 2.5 bar level. On a global scale, \citet{Momary07} report a pronounced asymmetry in the hemispherical flux emitted by Saturn. The northern hemisphere emits nearly twice the flux of the southern hemisphere, possibly indicative of strong seasonal changes in the formation and dissipation of aerosols in the upper troposphere ($\sim$300 mbar) of Saturn.

For this paper, we have undertaken an analysis of VIMS data of the Saturnian atmosphere from \emph{Cassini} up to 31 March 2007. We present descriptions of the basic morphology of the features seen in the VIMS imagery, and provide statistics on the latitudinal and size distribution of these features. We also perform a comparative analysis of the features seen in VIMS and ISS mosaics. We have adapted an automated cloud feature tracking algorithm to construct the zonal wind profile of Saturn's atmosphere from VIMS data. We compare this profile to previously derived profiles to estimate the vertical wind shear. 

All latitudes in this paper, unless explicitly stated otherwise, are planetographic.

\section{Data and Methodology}

\subsection{Data Sets and Common Procedures}
We obtained VIMS datacubes from the NASA PDS Imaging Node, and restricted our analysis to data sets of Saturn's atmosphere that either (i) attempted feature tracking by repeatedly observing the same absolute frame relative to System III longitude or (ii) monitored the overall state of Saturn's atmosphere via full-disk mosaics over a Saturnian day. A table listing all of the data that we analyzed is included as Table \ref{Table: vims_data}. We perform the standard VIMS data processing pipeline (e.g. removing cosmic ray hits, dividing by a pre-launch flat field, dividing by a solar spectrum, etc.) using software provided by the VIMS team. \citet{Barnes07} discusses this pipeline further. We used the ISIS (Integrated Software for Imagers and Spectrometers) software suite \citep{Gaddis97} for further data reduction and processing, which we discuss in the following paragraphs.

All of the VIMS datacubes in this study were prepared for analysis by first co-adding the five longest wavelength  channels in the data cube (5.057-5.122 $\mu$m) and omitting the remainder. With ISIS, we re-projected the image into a rectangular (simple cylindrical) projection, and used kernels provided by the Navigational and Ancillary Information Facility (NAIF) at the Jet Propulsion Laboratory for absolute image navigation and orientation. Our re-projected images that were used for both mosaic construction and feature tracking are oversampled at 0.1$^{\circ}$ pixel$^{-1}$ by utilizing a bicubic interpolation scheme supplied with ISIS. (1$^{\circ}$ on Saturn is $\sim$1000 km.) Typical VIMS images have a native resolution of about 0.25$^{\circ}$ pixel$^{-1}$ on Saturn, though this resolution strongly depends on the observation geometry and range from the target. Typical spacecraft ranges from Saturn during observations were $\sim$9 x 10$^5$ km, though these ranges sometimes varied by as much as 50\%.

A primary source of uncertainty in this paper is errors in the pointing knowledge of the spacecraft. Though the kernels provided by NAIF are carefully crafted, some uncertainty in the absolute location of where the spacecraft is pointing and the geographical location of the target is inevitable. We were unable to perform any manual correction to the pointing knowledge of the images via limb-fitting or other methods during the processing pipeline. We estimate the uncertainty in the absolute location of the cloud features to be 1--2$^{\circ}$ (J. Barnes, pers. comm.). However, the relative uncertainty in location between pixels is much smaller. Thus, our measurements of lengths and velocities, which are more dependent on relative uncertainty, do not degrade much in quality because of the small, but non-zero, absolute uncertainty.

\subsection{Mosaics}

We created global-scale mosaics of Saturn's atmosphere using VIMS data in order to characterize the frequency and size distribution of observed cloud features. In order to enhance the visible features and maximize the data return from the VIMS instrument, we implemented an unsharp mask procedure described in \citet{LeMouelic07}. This technique simply subtracts a percentage of a low-pass-filtered version of the original image from the original data. The resulting mosaics offer enhanced contrast and improved visibility for the backlit features. No special processing was performed to mitigate image seams.

We constructed two near-complete mosaics of Saturn's southern hemisphere, as \emph{Cassini} observed the southern hemisphere twice over a period of nearly two weeks in December 2006. During both observation sequences, the spacecraft acquired data by imaging the hemisphere in a 3x3 grid (with one corner absent) at regular intervals during a Saturnian rotation. The entire southern hemisphere is observed up to 80$^{\circ}$S, though some gaps in the coverage are present. We also constructed one mosaic of the northern hemisphere up to 60$^{\circ}$N from observations in September 2006. This mosaic has relatively lower spatial resolution from unfavorable orbital and observational geometry. Other observational sequences that covered the northern hemisphere with improved resolution during the \emph{Cassini} mission in the publicly available datasets were insufficient in global coverage or otherwise had unfavorable viewing conditions.

\subsection{Cloud Feature Census}

For the mosaics presented in this paper, we performed a statistical assessment of the population and characteristics of the features seen in Saturn's atmosphere. We limit our analysis to spot-like cloud features that are dark, compact, and ellipsoidal. These features are ubiquitous in the mid- and higher latitudes of the atmosphere. These features may be vortices, though we cannot explicitly confirm this because we currently are incapable of measuring local wind velocities within these structures. Furthermore, there are many complex and ambiguous structures where spots appear to be merging, shearing apart, or otherwise deforming. We elect to exclude these structures in our analysis in order to establish a lower bound in cloud feature population. To assess the statistics of the cloud features seen in 5 $\mu$m, we measured the central locations of these features to determine their latitudinal distribution. To measure lengths, we used the measuring tool in the ISIS software package to measure both the longitudinal and latitudinal span of these spots. This software tool is capable of precise, sub-pixel measurements, but we must be careful with their interpretation as we are examining over-sampled and interpolated mosaics. We estimate the uncertainty in the measurement of feature lengths to be at the level of 1 native resolution pixel ($\sim$250 km).  

\subsection{Automated Feature Tracking}

We adapted our automated feature tracker \citep{Choi07} to measure the zonal wind profile of Saturn's atmosphere from VIMS cubes. Here, cloud features are assumed to act as passive tracers of the ambient zonal winds. However, instead of using hand and eye to track features in images separated by a discrete time, we use software to extract portions of an image and compare them to nearby portions of a later image. Our software then determines which portions have the highest correlation and measures their offset, which is then used to determine wind velocity. This technique has been used to measure winds of Jovian vortices [\citet{Vasavada98}; \citet{Read05}; \citet{Choi07}], eddy velocity fluxes on Jupiter and Saturn [\citet{Salyk06}; \citet{DelGenio07}], and Venusian atmospheric velocities \citep{Toigo94}.

We examined five particular data sets (Table \ref{Table: vims_data}) out of more than twenty observation sequences that were designed for feature tracking. Unfortunately, we could not integrate the remaining data into our analysis due to poor observation geometry, map reprojection issues, poor image resolution, or a combination of all three. Our feature tracker normally attempts to measure velocities both in the zonal and meridional directions. However, we find no evidence of synoptic or regional scale meridional motion. Furthermore, the low spatial resolution of the VIMS data is insufficient for tracking local meridional motion. Thus, we modified our feature tracker to only measure zonal velocities. Subtle meridional shifts are apparent when manually blinking some images separated in time but are likely caused by uncertainties in camera pointing and image navigation. In addition, we adapted our feature tracker for the VIMS data by changing the size and shape of the correlation boxes that are extracted from the images. Small squares (the mode that our software typically uses) for correlation portions were unfit for the VIMS data as typical tracked cloud features were relatively large (a result of the low image resolution), and chances of returning a spurious result by the algorithm were high. We instead used correlation boxes that spanned a narrow range of latitudes ($\sim$1$^{\circ}$) but the full range of longitudes available in the image. The technique returns a single zonal velocity for each latitude. The precise latitudinal location of each velocity vector is set at the mid-point latitude of each rectangular correlation box. \citet{Limaye86} used a similar technique to construct one of the first zonal wind profiles of Jupiter. Typical image separation times for VIMS cube frames that we analyzed range from 10 to 30 minutes.

The 1$^{\circ}$ latitude correlation box leads to a $\sim$0.5$^{\circ}$ uncertainty for the vector's location, since the vector could result from feature(s) anywhere within the box. We can estimate the effect that the locational uncertainty has on the velocity measurement by multiplying it by the magnitude of latitudinal wind shear present. We estimate that the typical uncertainty is $\sim$10 m s$^{-1}$, though it could be up to 30--40 m s$^{-1}$ in a high latitudinal shear environment. However, this is an upper bound to the uncertainty, as it is unlikely that the tracked features in a correlation box are all located at one edge of the box. Overall, the uncertainty in the zonal wind measurements is insufficient to significantly affect our conclusions.

\section{Results}

\subsection{Atmospheric Features}
\label{Section: Atmospheric Features}

Much like the southern hemisphere, near-infrared continuum filter mosaics composed from \emph{Cassini}'s Imaging Science Subsystem (ISS) \citep{Porco04} analyzed by \citet{Vasavada06}, the VIMS mosaics (Figures \ref{Figure: mosaic_southhem1}--\ref{Figure: mosaic_northhem}) exhibit a gradation with latitude in cloud feature clarity: the diffuse and muddled features in the lower latitudes contrast sharply with the more well-defined, distinct features seen in higher latitudes. The relative lack of well-defined, high-contrast features in the equatorial region supports the presence of an optically thick haze in the upper atmosphere capable of attenuating the thermal emission. However, occasional dark, wispy cloud features are visible in the equatorial region (Figure \ref{Figure: eq_feat}). These are likely clouds at deeper altitudes that perhaps erupt periodically and are more effective at blocking the thermal emission. Equatorial 5 $\mu$m hotspots similar to those encircling Jupiter (\citet{Ortiz98} and \citet{Arregi06}) have not been observed at low-latitudes, a result noted previously by \citet{Yanamandra01} through ground-based telescopic observations.

The overall cloud morphologies in the VIMS and ISS mosaics are also broadly similar. The tropical latitudes in both hemispheres are dominated by numerous tilted, alternating dark and light linear features that strongly resemble the ``tilted cloud streaks" noted by \citet{Vasavada06} in their study. An especially strong wind shear environment (10 to 20 m s$^{-1}$ per degree of latitude) is present at these latitudes. Interspersed between the tilted features are zonally aligned light and dark stripes located at approximately 16$^{\circ}$S, 22$^{\circ}$S, 28$^{\circ}$S, and 15$^{\circ}$N. These bands span a wide range of longitudes but appear interrupted at some places and are not as distinct as the bands seen at higher latitudes. The region between 30$^{\circ}$S--35$^{\circ}$S is a transitional region containing both occasional spot features (including a few with spiral patterns surrounding them, suggesting circulation) and stripes (both light and dark) occasionally interrupted by ambient clouds. A pair of narrowly separated zonal jets at 31$^{\circ}$S and 33$^{\circ}$S are perhaps responsible for the complex meteorological patterns. A similar transition region containing two light stripes with ragged cloud features interspersed in between them is present between 30$^{\circ}$N and 35$^{\circ}$N. However, no known zonal jet exists at this latitude in the northern hemisphere.

The mid-latitudes (poleward of 35$^{\circ}$ latitude) in both hemispheres of Saturn are notable for the presence of numerous compact dark spot features, indicating regions of high cloud opacity from their efficiency at blocking Saturn's thermal radiation. The majority of these spots are singular and isolated, though many instances of complex structures (apparent mergers, possible interacting structures, and amorphous shapes) are present interspersed among the singular spots. As discussed in Section \ref{Section: Spot Feature Statistics}, these dark features are not distributed evenly across latitudes but instead concentrate in dense populations between distinct, zonally-aligned stripes. These stripes correspond to locations of zonal jets (discussed further in Section \ref{Section: Zonal Wind Profile}), unlike the subtle bands seen in tropical latitudes, which are not associated with the jets. The image resolution and temporal coverage of the observations are insufficient for estimating the vorticity of these features, and we cannot definitively state how many of these spots are vortices. 

A salient feature of the northern hemisphere in the VIMS mosaics is the so-called ``string of pearls'' that exists in a narrow band just south of 40$^{\circ}$N. These features are areas of uncharacteristically high thermal emission, indicative of low cloud opacity. We catalogued 22 ``pearls" in the northern hemisphere mosaic, and examination of the central locations of these pearls reveals that they roughly divide into three groups. The first group of five and the second group of ten have similar spacings between pearls of about 3.7$^{\circ}$ longitude on average. However, the third group of seven are each spaced slightly wider, with a mean separation of approximately 4.8$^{\circ}$. The shapes of the pearls are slightly elliptical, with a median aspect ratio (defined as east-west length divided by north-south length) of approximately 1.1. However, we note that there are some examples of pearls with shapes at both extremes, with aspect ratios ranging from 0.5 to 1.7. 

Two possible analogues currently exist for these features. Our preferred hypothesis is that the pearls are a manifestation of a von K\'arm\'an vortex street, a repeating pattern of vortices where cyclones and anticyclones alternate with one another. The contrasting regions of cloud opacities could correspond to alternating regions of relative vorticity, with 5 $\mu$m bright regions being cyclonic and dark regions being anticyclonic. The proximity of the pearls to 40$^{\circ}$N, a quasi-westward flowing band wedged between strongly eastward flowing jets (see Figure \ref{Figure: vims_zprof}) supports the vortex street model. The White Ovals, a system of anticyclones thought to be part of a von K\'arm\'an vortex street on Jupiter \citep{Youssef03}, were located in a similar position in Jupiter's zonal wind profile (a westward flowing band in between two eastward jet streams). We again caution that we cannot directly measure the local wind flow of these features and thus cannot confirm that they are vortices. However, preliminary numerical simulations of Saturn's mid-latitude atmosphere support our hypothesis of a vortex street model for the pearls (K. Sayanagi, pers. comm.). An alternative possibility is that these areas of enhanced emission result from cloud modulation caused by wave motion, similar to the 5 $\mu$m equatorial hotspots on Jupiter, though the ``pearls" appear to be more numerous and smaller than their Jovian counterparts. Whereas equatorial hotspots on Jupiter range in length from 5,000 to 10,000 km \citep{Orton98}, the average length of a pearl on Saturn is just over 1,000 km. Another contrast between the two features is the observation that there are typically $\sim$10 Jovian 5 $\mu$m hotspots that globally span the latitude band between 5$^{\circ}$N and 10$^{\circ}$N \citep{Arregi06}. In contrast, Saturnian pearls appear to be confined to a span of longitude approximately 100$^{\circ}$ in width, and are more numerous. The leading explanation for Jovian hotspots is that they are a manifestation of the downwelling branch of an equatorially trapped Rossby wave [\citet{Allison90}; \citet{Showman00}; \citet{Friedson05}]. This physical model may also extend to the Saturnian pearls, though the wave would presumably not be equatorially trapped. A hybrid model with a Rossby wave acting to trap vortices and preventing them from merging is also possible. \citet{Youssef03} have suggested such a model in the case of Jupiter's White Ovals throughout much of the 20th century. 

\subsection{Zonal Wind Profile}
\label{Section: Zonal Wind Profile}

Figure \ref{Figure: vims_zprof} is our constructed zonal wind profile. For comparison, we have also included the zonal wind profiles constructed from \emph{Voyager} \citep{Sanchez-Lavega00} and \emph{Cassini} \citep{Vasavada06} imaging data. Generally, our zonal wind profile is in excellent agreement with the shape of previous zonal wind profiles from \emph{Voyager} and \emph{Cassini} and with previous preliminary reports about the VIMS dataset \citep{Baines05DPS}, indicating that the overall jet structure has remained broadly consistent. An additional work by \citet{Baines05EMP} states a measurement of 390 $\pm$ 50 m s$^{-1}$ for a feature at 8$^{\circ}$S, which is in agreement with our results. The broad equatorial jet is clearly evident, and its latitudinal shear is also in reasonably good agreement with previous zonal wind profiles. Our analysis of the VIMS images provides evidence for rapid equatorial flow at the jet core: clouds just north of the equator flow at $\sim$500 m s$^{-1}$, and perhaps even faster. This suggests that the near factor-of-two reduction in speed relative to \emph{Voyager} measurements for the equatorial jet \citep{Sanchez-Lavega03} and the more modest reduction measured by \citet{Porco05} could perhaps be confined to the upper troposphere and that VIMS may be sensing deeper levels of the jet at these latitudes. We express caution with these particular measurements of rapid flow at the jet core (i.e. the cluster of points with $u >$ 500 m s$^{-1}$ at $\sim$1--5$^{\circ}$N and 7$^{\circ}$S), as the equatorial region typically lacks discrete, high-contrast cloud features that are easily trackable by our technique. However, we have performed manual cloud tracking of a small cloud complex embedded in the flow (seen earlier as Figure \ref{Figure: eq_feat}) that support the measurements above 500 m s$^{-1}$. We note that it is possible that the highest-speed, tracked cloud features are a local disturbance and not characteristic of the zonally-averaged flow. Regardless, it seems clear that the equatorial jet speed exceeds $\sim$400--450 m s$^{-1}$ at the pressure sensed by VIMS 5 $\mu$m images.

A few higher-latitude jets are present in the VIMS data, though unfortunately we cannot derive results south of 60$^{\circ}$S using our current techniques on the publicly available data set. There is remarkably good agreement among all three zonal wind profiles for the precise latitudinal location of the jets. However, differences exist in the peak speeds measured at the jet cores, with the exception of the jet near 45$^{\circ}$N, where our derived speeds from VIMS is nearly identical to that from \emph{Voyager}. Furthermore, there is no universal increase or decrease of the peak velocities across all jets between the VIMS profile and past profiles. The jet cores at 75$^{\circ}$N and 50$^{\circ}$S appear to be flowing faster than in the previous profiles. The measurement at 75$^{\circ}$N is attributed to a cluster of clouds located north of a large dark spot feature, and these clouds appear to be embedded in a relatively wide westward-flowing jet. At 50$^{\circ}$S, a cloud feature that is embedded in a dark stripe (presumably corresponding to a narrow jet stream) is the likely source for our measurement. Though manual inspection of these features support our automated measurements, it is important to note that it is unclear how closely the motion of these features represent the ambient zonally-averaged zonal flow. The measurement at 60$^{\circ}$S also supports the presence of apparent wind shear compared with past profiles, though we feel this measurement is less convincing from lack of repeat observations. One of the starkest differences between the VIMS zonal wind profile and previous profiles is the jet at 65$^{\circ}$N, which appears to be flowing at nearly half of the speed of the jet at that latitude as seen by \emph{Voyager}. Upon further manual inspection of the feature tracking images at that latitude, the evidence for wind shear at this location is inconclusive because of the low image resolution. 

\subsection{Spot Feature Statistics}
\label{Section: Spot Feature Statistics}

Our database of spot features consists of 578 and 572 entries for the two southern hemisphere mosaics, and 297 entries for the northern hemisphere. For the purposes of comparison, we have also analyzed the southern-hemisphere ISS mosaic composed and examined by \citet{Vasavada06}, their Figure 1. This mosaic is composed of images observed using one of the near-infrared (750 nm) continuum filters. We only analyze the compact, patchy, light spots (clouds) in the ISS mosaic, as \citet{Vasavada06} have already reported their analysis on the darker vortices. (It is important to note here that features that appear dark in VIMS 5 $\mu$m images are simply more effective at blocking the thermal emission, whereas features that appear dark in ISS images are inherently lower in albedo.) We catalogued 687 light spots in the ISS mosaic, and used the same ISIS analysis tools throughout all four mosaics. 

Figure \ref{Figure: hist_lat} is a histogram showing the number of spot features as a function of their latitude for all four of the mosaics that we have analyzed. We sorted the features into latitude bins 1$^{\circ}$ in size, centered at each integer latitude. The histograms for the southern hemisphere VIMS mosaics (Fig. \ref{Figure: hist_lat}a and \ref{Figure: hist_lat}b) demonstrate that the features are primarily located in three bands that are demarcated by zonal jet streams. The latitudinal distribution of these features is remarkably similar in the two mosaics, with minor differences between the two mosaics attributable to irregularities in observational coverage and inherent variability in the populations and lifetimes of these features. A fourth band containing spot features is suggested at higher latitudes ($> 75^{\circ}$S). However, because of the poor image resolution and the cutoff in the image beyond 80$^{\circ}$S, we are probably under-representing the population of features in this latitude band.

Figure \ref{Figure: hist_lat}c is the number distribution of the light, high-albedo spots in the ISS mosaic as a function of latitude. For comparison, we have included the population of the dark, low-albedo spot features counted by \citet{Vasavada06} as black segments to supplement our results. The features are clearly divided into four latitudinal zones largely similar to the distribution seen in our southern hemisphere VIMS mosaics, with some differences in the number of features seen in each latitude zone between the ISS and VIMS mosaics. South of 50$^{\circ}$S, we detect more features in the ISS mosaics than in the VIMS mosaics, with substantially more in the southern polar regions. This is likely a result of improved image resolution and facilitated feature identification at the higher latitudes in the ISS images (rather than incomplete longitudinal coverage, as we will show in Figure \ref{Figure: hist_num_nrml_vs_lat}). 

Figure \ref{Figure: hist_lat}d is the number distribution of dark spots in the northern hemisphere VIMS mosaics. The tropical latitudes contain occasional spots, with the numbers increasing gradually with latitude. A main group of spots is centered poleward of 40$^{\circ}$N, with a secondary group of spots present spanning 52-59$^{\circ}$N. A broad gap in the spot population at 45-50$^{\circ}$N aligns with a strong prograde jet at that latitude band. This matches the basic structure seen in the southern hemisphere, where zonal jets mark latitudes with a deficiency of spots. However, this structure does not extend to the tropical latitudes, as gaps in the population distribution exist on the flank of the fast equatorial jet.

The histograms shown in Figure \ref{Figure: hist_lat} are simply the number of spots that occur in latitude bins 1 $^{\circ}$ wide, as a function of latitude, in our mosaics. However, our longitudinal coverage varies from latitude to latitude; moreover, the circumference of any given latitude circle decreases with increasing latitude. It is thus useful to normalize the spot numbers to remove these effects.  To do so, we calculate the number of spots $\overline{N}$ that occur in latitude bins 1 $^{\circ}$ wide per unit longitudinal distance. This is given by

$$\overline{N} = \frac{N}{2\pi fR_{curv}},$$

where $N$ is the number of spots in a latitude bin, $f$ is the fraction of mosaic coverage at that latitude (unitless), and $R_{curv}$ is the zonal radius of curvature. Figure \ref{Figure: hist_num_nrml_vs_lat} shows this quantity for all four mosaics. Because $\overline{N}$ compensates for the gaps in the VIMS mosaic coverages, any remaining differences between Figures \ref{Figure: hist_num_nrml_vs_lat}a and b with \ref{Figure: hist_num_nrml_vs_lat}c suggest that other factors account for the discrepancies in the spot counts. The larger normalized count in the ISS mosaic at high latitudes (poleward of 60$^{\circ}$S) compared to the VIMS counts in these bands is likely caused by improved image resolution in these areas. The smaller normalized count in the ISS mosaic at mid-latitudes (37$^{\circ}$S-48$^{\circ}$S) is more puzzling. This could be indicative of increased activity or a more complex vertical structure confined to these latitudes. 

Figure \ref{Figure: hist_ewdist} presents a number-size distribution for all four mosaics, including the spot feature population in \citet{Vasavada06}. The overwhelming majority of the spot features are below 1500 km in east-west diameter. However, a few features have diameters that exceed 2000 km, with one feature in both southern hemisphere VIMS mosaics exceeding 5000 km diameter. The ISS southern hemisphere distribution (Fig. \ref{Figure: hist_ewdist}c) reveals a slight difference in the location of the peak size. Whereas the VIMS southern hemisphere size distribution peaks at about 600 km, the ISS distribution peaks at a lower size, about 400 km. This may be an artifact of enhanced resolution in the ISS mosaics and as a result, improved detectability of the smallest features. The peak in the size distribution at 400-600 km has implications for Saturn's Rossby radius of deformation, a length scale at which rotational (Coriolis) effects become comparable to buoyancy effects. One can expect vortices at a size near the deformation radius, as structures that develop from inverse energy cascade and geostrophic adjustment typically attain sizes near this length scale [\citet{Cho96}; \citet{Polvani94}]. If we assume that the majority of this population are small vortices, it suggests that the deformation radius is within a factor of two of $\sim$500 km. This estimate is relatively small compared to estimates of the deformation radius on Jupiter (\citet{Showman07} reports 1000-3000 km based on his numerical simulations) and in the upper troposphere of Saturn (2000-8000 km from \citet{Read07} via an analysis of the zonal wind, temperature, and molecular hydrogen para-fraction profile).  

We examine a feature's east-west diameter as a function of its latitude in Figure \ref{Figure: distvslat} and Table \ref{Table: diam_stats}. Clearly, most of the spot features are below 1500 km in diameter, with few features above this threshold. We note a subtle decrease in the characteristic diameters of the features with increasing latitude in both hemispheres. This slight trend also suggests that the deformation radius is a controlling factor in the formation of these spot features. The deformation radius is inversely correlated with the Coriolis parameter $f$; therefore, the radius will decrease with higher latitude.

Figure \ref{Figure: aspectratio} presents the aspect ratios for all of the features catalogued in our study. We define aspect ratio as the ratio of a feature's principal east-west diameter to its principal north-south diameter. Most of the features span a broad range of aspect ratios, as there is no obvious bias towards a particular shape for these features in the main population below 1500 km diameter. However, we note that larger features ($>$ 2000 km) exhibit a preference for an aspect ratio greater than unity (zonal elongation). Aspect ratios for visible spots have been measured before on Jupiter and Saturn. \citet{Maclow86} noted that aspect ratios tended to be greater than unity with increasing east-west diameters based on their analysis of spots on Jupiter from \emph{Voyager} imagery. \citet{Li04} demonstrate the same trend based on their analysis of Jovian spots from \emph{Cassini} imagery. Both studies reveal that spots on Jupiter with diameters above 3000 km tend to be more zonally elongated. Saturn appears to also follow this general rule. \citet{Vasavada06} also show that the dark, low-albedo vortices on Saturn tend to have aspect ratios greater than unity, though we note that from both Saturn studies, there is a general deficiency in the feature population above 3000 km compared with the population found in Jupiter's atmosphere. This property of the features in both Jupiter's and Saturn's atmospheres may be a manifestation of the $\beta$-effect (the variation of the Coriolis force with latitude) acting to preferentially elongate structures that exceed a diameter of $\sim$2000 km.

\section{Discussion}
\label{Section: Discussion}

\citet{Bjoraker07} estimates that the 5 $\mu$m thermal radiation emanates from a pressure level of 5 bars. However, the depth at which the thermal radiation originates does not necessarily have to coincide with the depth of the imaged features in the VIMS mosaics. Preliminary radiative transfer analysis by \citet{Baines05DPS} and \citet{Momary06} indicate that the features observed by VIMS are near the 2-bar level. Utilizing information gleamed from the ISS and VIMS zonal wind profiles, assuming that they originate from different depths, yields constraints about the latitudinal temperature gradient of Saturn's atmosphere. Saturn's winds are in geostrophic balance, and vertical shear of geostrophically balanced winds is controlled by the thermal wind equation \citep{Holton04} 

$$-\frac{\partial \mathbf{v}_g}{\partial p} = \frac{R}{pf} \mathbf{\hat{k}} \times \nabla_p T,$$

where $\mathbf{v}_g$ is the geostrophic wind, $p$ is pressure, $R$ is the specific gas constant, $f$ is the Coriolis parameter, and $T$ is the temperature. Thus, with all other factors being equal, vertical shear in the geostrophic zonal winds is directly correlated with the latitudinal gradient of temperature. If we assume that the features tracked by ISS and VIMS are separated by a scale height, we can place a bound in the vertical wind shear using the uncertainty in the zonal wind profile. Using our uncertainty estimate of $\pm$10 m s$^{-1}$, $\partial{T}/\partial{y}$ has an upper bound of $\sim$0.5 K/1000 km. When extrapolated over 5,000--10,000 km, our zonal wind profile rules out latitudinal temperature contrasts exceeding 2--5 K throughout most of the atmosphere. 

However, apparent wind shear is present at certain jet stream cores. Our zonal wind profile shows faster wind speeds at the equator and at the 48$^{\circ}$S jet core compared with previous profiles. Recently, \citet{Sanchez-Lavega07} used methane and near-infrared filter images from \emph{Cassini} ISS, and observed higher-altitude features flowing at a slower speed compared to features at the main cloud deck ($\sim$700 mbar), yielding a vertical wind shear $\partial u/\partial z \sim$ 40 m s$^{-1}$ per atmospheric scale height $H$ ($\sim$50 km at the equator). \citet{Garcia-Melendo08} has extended that work to show that vertical wind shear (estimated at 10 m s$^{-1}$ per atmospheric scale height $H$) is present at several southern hemisphere zonal jets. However, if we again assume that the features tracked by ISS and VIMS are separated by a scale height, the $\sim$100 m s$^{-1}$ difference at the jet cores implies a shear that is greater than the estimate from \citet{Sanchez-Lavega07} by a factor of 2--3. (Our wind shear is also well above the $\sim$25 m s$^{-1}$ per atmospheric scale height at the 500 mbar level estimated by \citet{Flasar05}.) Importantly, our VIMS cloud tracking, coupled with ISS cloud tracking results, allows us to constrain the latitudinal temperature gradients at pressures significantly deeper than that directly measurable from thermal-emission spectra with, for example, \emph{Cassini}'s composite infrared spectrometer (CIRS) \citep{Flasar04} instrument. Our measured vertical wind shear implies a local latitudinal temperature gradient at the 2 bar level to be $\sim$1 K/1000 km at the equator, and as high as $\sim$5 K/1000 km at the 48$^{\circ}$S jet. Our estimates are upper bounds, as shear would be lower if VIMS were observing features in a deeper ($>$ 2 bar) cloud deck. Thus, our analysis implies that the latitudinal temperature gradients at specific latitudes may exceed gradients found at other latitudes at the same pressure. 

\begin{figure}[htp]
  \centering
  \includegraphics[width=6.5in, keepaspectratio=true]{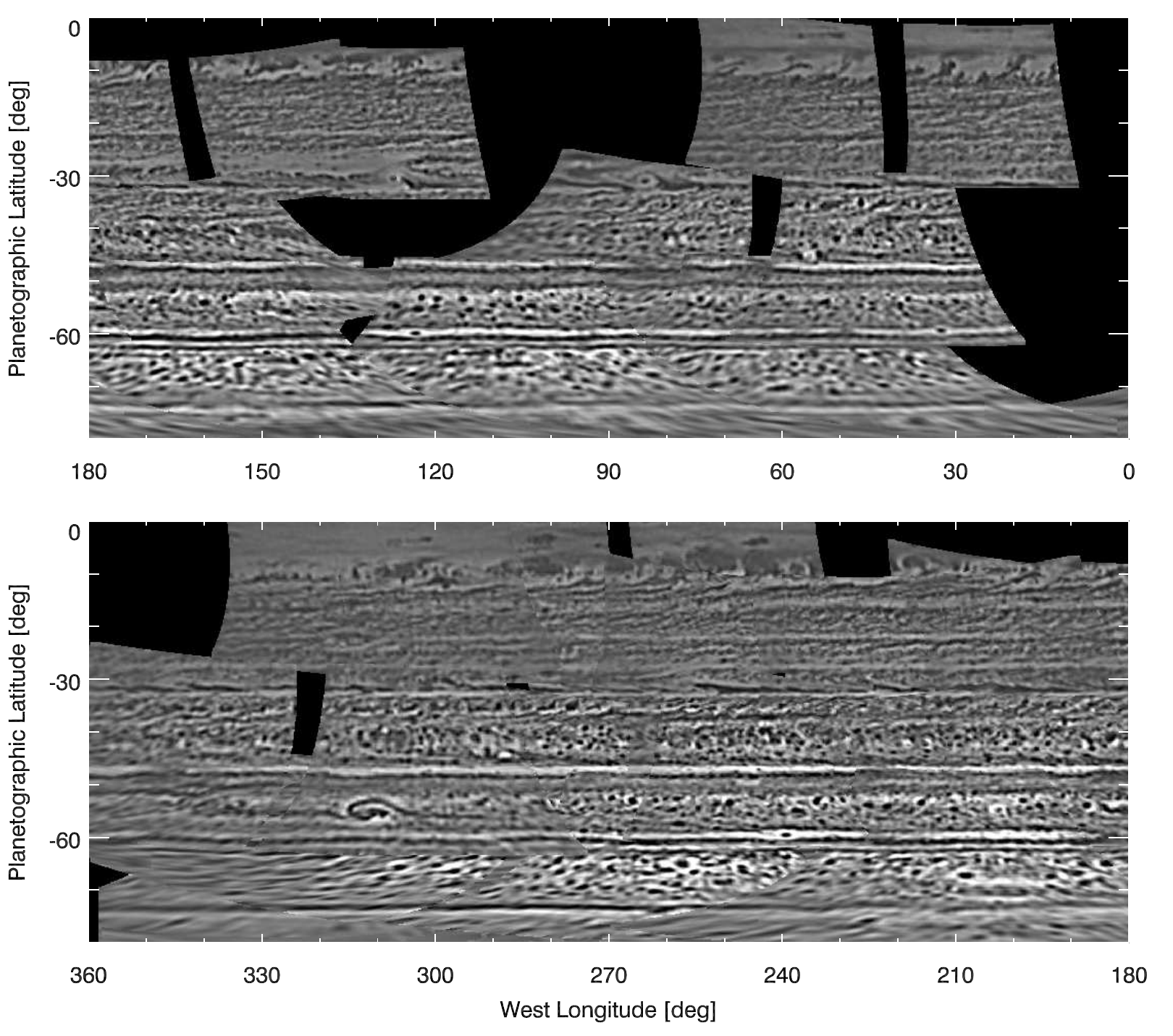}
  \caption[Saturn's Southern Hemisphere as imaged by VIMS, Dec. 1 2006]{
    \label{Figure: mosaic_southhem1}
    \textbf{VIMS Southern Hemisphere Mosaic I.} A snapshot of Saturn's southern hemisphere as observed in 5 $\mu$m by VIMS. This mosaic is an assembly of images taken on 1 December 2006. Each frame component of the image has been projected in a cylindrical (rectangular) projection before final construction of the mosaic.}
\end{figure}

\begin{figure}[htp]
  \centering
  \includegraphics[width=6.5in, keepaspectratio=true]{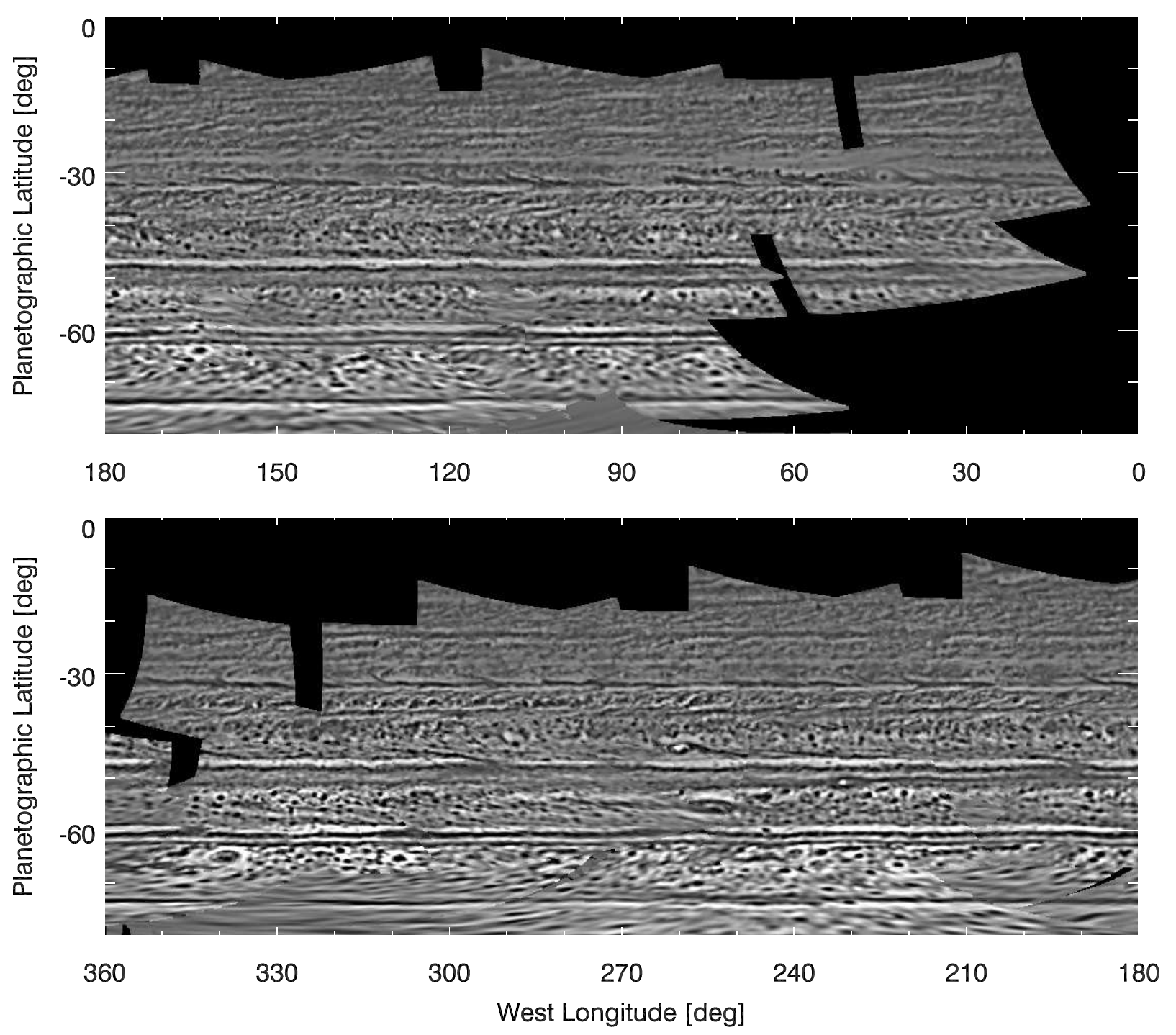}
  \caption[Saturn's Southern Hemisphere as imaged by VIMS, Dec. 13 2006]{
    \label{Figure: mosaic_southhem2}
    \textbf{VIMS Southern Hemisphere Mosaic II.}
    Same as Figure \ref{Figure: mosaic_southhem1}, but approximately two weeks later.}
\end{figure}

\begin{figure}[htp]
  \centering
  \includegraphics[width=6.5in, keepaspectratio=true]{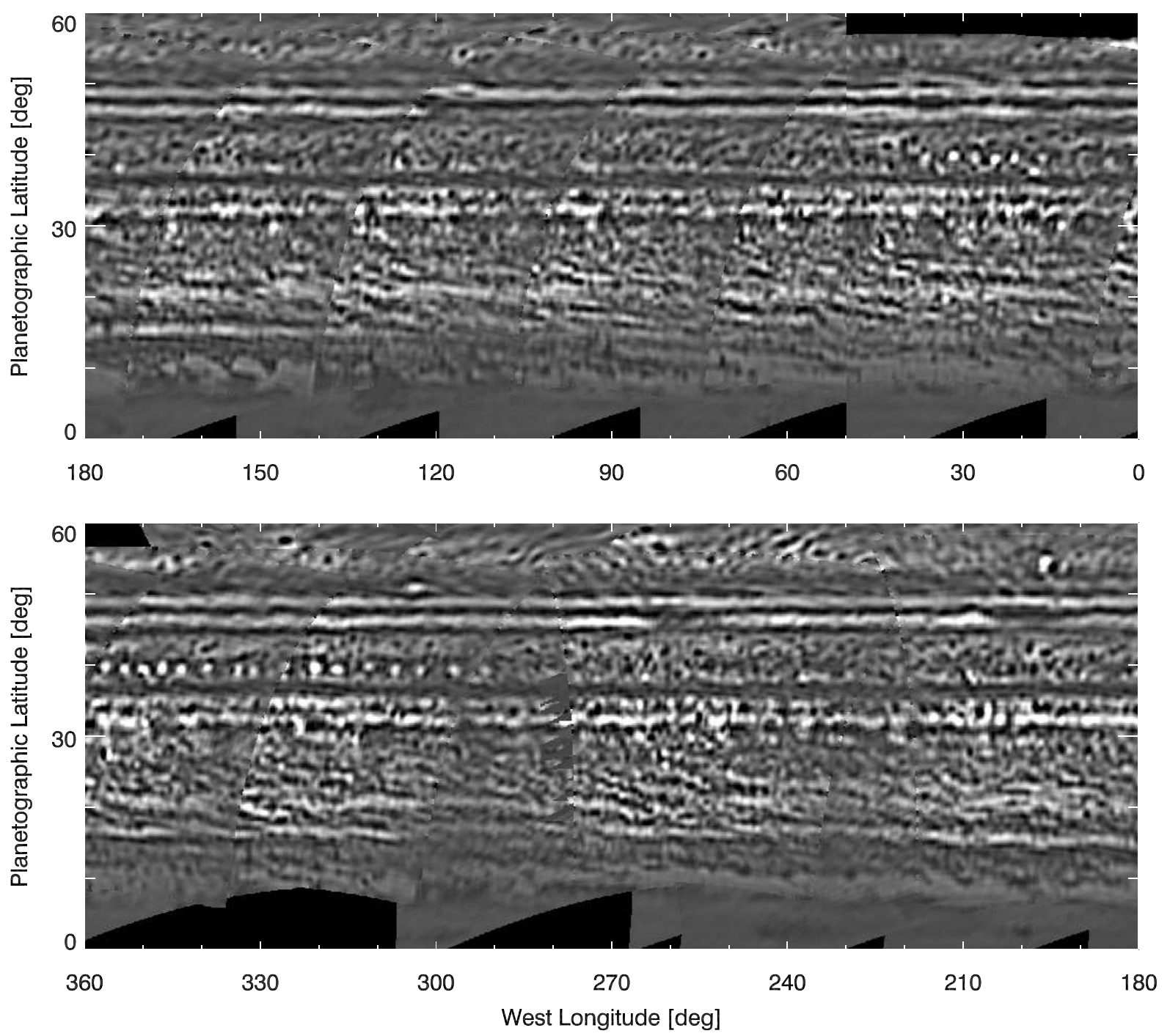}
  \caption[Saturn's Northern Hemisphere as imaged by VIMS, 2006]{
    \label{Figure: mosaic_northhem}
    \textbf{VIMS Northern Hemisphere Mosaic.}
    Saturn's northern hemisphere as seen by VIMS in the 5-micron spectral window. This mosaic is an assembly of images taken in April 2006.}
\end{figure}

From our analysis of the cloud features in the VIMS and ISS mosaics, it appears plausible that VIMS is simply observing the same cloud features that have been observed by visible imagers from \emph{Cassini} and \emph{Voyager}. Figure \ref{Figure: vims_and_iss} supports our hypothesis. On the left of Figure \ref{Figure: vims_and_iss} is a portion of an ISS mosaic analyzed by \citet{Vasavada06}, whereas the right portion of Figure \ref{Figure: vims_and_iss} is an \emph{inverted} portion of Figure \ref{Figure: mosaic_southhem1}, one of the southern hemisphere VIMS mosaics. Note the general similarities between both mosaics in the alignment of the horizontal stripes, the bright, white cloud features at higher latitudes, and the possible long-lived vortex just north of 60$^{\circ}$S. The appearance of the stripes at higher latitudes also match well, indicating that the high-albedo, white bands (suggestive of a thick cloud band) observed in ISS images are dark in the 5 $\mu$m images (indicating high opacity), and dark bands are bright in the 5 $\mu$m data set (indicating low opacity).

Comparison of the statistics of the spotted features in the VIMS and ISS mosaics also suggests that both instruments are typically observing the same features. Similarities in the features' latitudinal distribution and numerical population are evidence that the majority of the VIMS dark spots are white cloud patches seen in the ISS images. These bright cloudy areas likely contain abundant aerosols that are efficient at blocking abyssal thermal radiation. Unfortunately, this creates a degeneracy for the VIMS dark spots as both the patchy white clouds and the dark, low-albedo vortices contribute to the VIMS dark spot population. One key difference based on Figure \ref{Figure: hist_ewdist} is that features analyzed by \citet{Vasavada06} that exceed 1000 km diameter seem to be largely missing in the VIMS mosaics, as only a small percentage of VIMS features have diameters above this threshold. However, we note that based on Figure 8 of \citet{Vasavada06}, most of the dark vortices in the ISS images with diameters below 1000 km are simply classified as ``dark.'' In contrast, most of the ISS vortices with diameters above 1000 km are classified as ``dark with bright margin,'' ``bright-centered,'' or simply ``bright.''  Thus, it is possible that these more complex vortex families have a different structure when observed in 5 $\mu$m and were overlooked in our analysis. 

Overall, we believe that the majority of the dark spotted features seen in the VIMS mosaics are the same features as the ubiquitous white and dark spots seen in ISS images. In one scenario, ISS and VIMS could simply be observing the same cloud layer; lateral modulation of that cloud deck would then produce lateral variations in reflected sunlight (hence producing the ISS features) as well as lateral variations in 5-micron emission to space (hence producing the VIMS 5 $\mu$m features). However, estimates of the altitudes of features observed in ISS are $\sim$700 mbar, which is substantially less than that suggested for the depth of the VIMS features ($\sim$2 bars).  Alternatively, we propose a scenario where ISS is observing the reflected cloud tops of an ammonia cloud feature aloft, and VIMS is observing the base of this cloud feature at depth. This scenario implies that the features are nearly a scale height (or more) in thickness.  For this scenario to work, the upper-level cloud tops that cause reflection of short-wavelength sunlight (as observed by ISS) must have small-enough particle size so that they do not block the 5 $\mu$m radiation upwelling from deeper levels.  One thus needs a vertical gradient in particle size, with smaller particles aloft and larger particles at the base.  The cloud structure associated with a given feature (e.g., a vortex) could thus cause reflection of sunlight from pressures of $<$ 1 bar while allowing emission of 5 $\mu$m radiation from $\sim$2 bars. In any scenario, however, we must remain careful in interpreting the structure of Saturn's atmosphere at the jet streams, as VIMS may be observing tracers of motion housed in a deeper NH$_4$SH, or perhaps H$_2$O cloud deck.

Though we have not directly observed rotation within these spot features, we believe that these spots are vortices. Our analysis suggests that the deformation radius of Saturn's atmosphere controls both their inherent size and the variance of their size with latitude. Furthermore, \citet{Penny08} suggest that the Rhines effect \citep{Rhines75} is suppressed at the latitudes containing VIMS spot features, indicating that these latitudes can generate and support vortices. Our proposed hypothesis that the spots have finite thickness is also consistent with the dynamics of coherent vortices; the vertical extent of such a feature is typically $f/N$ times their horizontal width, where $f$ is the Coriolis parameter and $N$ is the Brunt-V\"ais\"al\"a frequency [\citet{Charney71}; \citet{Reinaud03}].  For Saturnian estimates (f $\sim$ 10$^{-4}$ s$^{-1}$ and N $\sim$ 0.001--0.005 s$^{-1}$), a vortex 1000 km across would then be 20-100 km or 0.5-3 scale heights thick. Moreover, numerical models of vortex evolution show success in matching vortex observations when the vortices are 1-3 scale heights thick \citep{Morales-Juberias05}. 

\section{Summary and Future Work}

We have performed an analysis of an extensive data set about the Saturnian atmosphere in the 5 $\mu$m spectral window from the VIMS instrument on board \emph{Cassini}. Our mosaics of both the northern and southern hemispheres reveal an extensive population of dark, compact spots that are areas of blocked thermal emission. A statistical analysis of these features and comparison with earlier visible-light mosaics of the southern hemisphere suggest that the majority of the ordinary spot features seen in the VIMS mosaics are white cloud patches, with a minority population of dark (in reflected sunlight) vortices. Automated cloud feature tracking reveals that the general structure of the zonal wind profile is largely similar to profiles constructed from \emph{Voyager} and \emph{Cassini} data. Differences in the speeds exist at some latitudes, indicative of longitudinal variability or observation of deeper features indicating vertical wind shear. 

Many opportunities are available for continued analysis of the rich VIMS dataset concerning Saturn's atmosphere. Further radiative transfer studies will help constrain the altitudes of the features observed by VIMS and refine our estimates of the vertical wind shear and temperature gradients in the atmosphere. Analysis of features at polar latitudes was beyond the scope of this study, but interesting features are located at these latitudes, including a polar vortex above the south pole and a hexagonal standing wave above the northern polar regions \citep{Baines07}. \emph{Cassini}'s extended mission will yield an opportunity to measure seasonal changes in the chemistry, meteorology, and structure of the atmosphere. Additional observation sequences designed for feature tracking will allow us to supplement our zonal wind profile and measure the degree of vertical wind shear throughout all latitudes. Furthermore, the majority of the VIMS data that we have analyzed are observations of the night side of Saturn, preventing us from comparing simultaneous VIMS and ISS images for common features. Comparative studies of the ISS and VIMS data sets and future planned observations will test vertical cloud structure models. We look forward to continued analysis of the data returned by VIMS and the \emph{Cassini} spacecraft as we continue to explore Saturn's unique atmosphere.

\acknowledgments
We thank our reviewers, Dr. Ashwin Vasavada and Dr. Andrew Ingersoll, for providing helpful comments that strengthened this paper. We thank Dyer Lytle and Virginia Pasek for their assistance in this project. We also thank Joe Plassmann and the staff of the Planetary Image Research Laboratory at the University of Arizona for their assistance and maintenance of some of the computing resources used for this research. We thank Dr. Ashwin Vasavada and Sarah H\"{o}rst for providing us with the ISS mosaic analyzed in this paper and their data.  

%
%
%
%
%
%
%
%
%
%




%
%

\begin{table}
\begin{center}
\begin{tabular}{lcc}
\hline
Data set & Latitude & Spacecraft Clock Time \\
\hline

\textbf{Feature Tracking}\hspace{2cm} & \hspace{1cm}15$^{\circ}$S--60$^{\circ}$S\hspace{1cm} & 1524828885--1524829781 \\
 & \hspace{1cm}29$^{\circ}$S--42$^{\circ}$S\hspace{1cm} & 1524977581--1524980093 \\
 & \hspace{1cm}3$^{\circ}$S--18$^{\circ}$S\hspace{1cm} & 1534393899--1534403233 \\
 & \hspace{1cm}4$^{\circ}$N--36$^{\circ}$N\hspace{1cm} & 1534575912--1534578687 \\
 & \hspace{1cm}2$^{\circ}$S--80$^{\circ}$N\hspace{1cm} & 1546355125--1546358126 \\
 \textbf{N. Hemisphere Mosaic I} & \hspace{1cm}0$^{\circ}$--60$^{\circ}$N\hspace{1cm} & 1536703543--1536741990 \\
\textbf{S. Hemisphere Mosaic I} & \hspace{1cm}0$^{\circ}$--80$^{\circ}$S\hspace{1cm} & 1543657108--1543694887 \\
\textbf{S. Hemisphere Mosaic II} & \hspace{1cm}5$^{\circ}$S--80$^{\circ}$S\hspace{1cm} & 1544708794--1544739494 \\
 
\end{tabular}
\caption{\label{Table: vims_data} VIMS data sets used in this study.}
\end{center}
\end{table}

\begin{figure}[htp]
  \centering
  \includegraphics[width=5.5in, keepaspectratio=true]{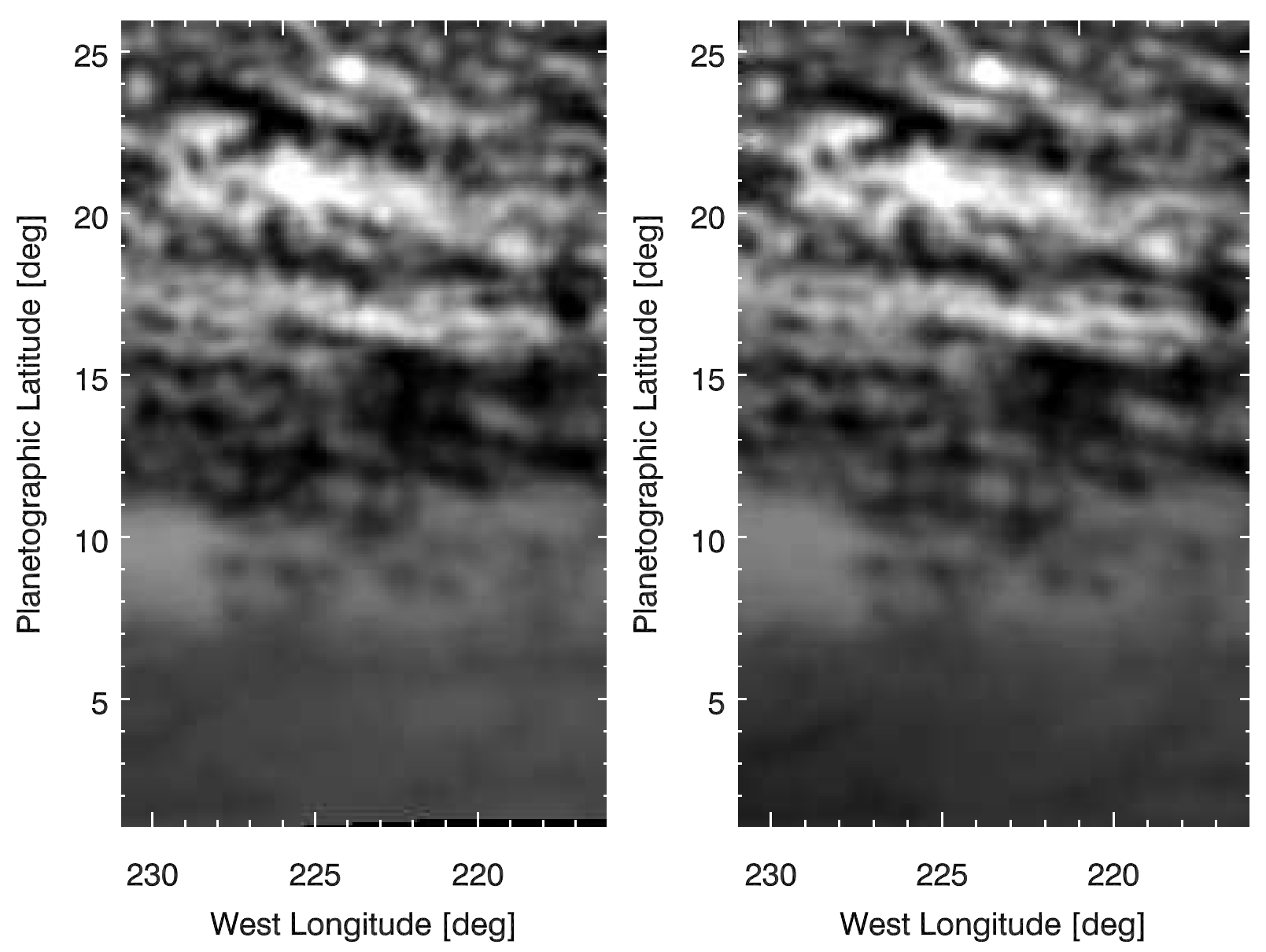}
  \caption[Equatorial Cloud Feature]{
    \label{Figure: eq_feat}
    Rectangular projection of VIMS data cubes in the 5-micron spectral window observing the northern tropical latitudes. The time separation of these two images is 29m 22s.}
\end{figure}
\ref{Figure: eq_feat}

\begin{figure}[htp]
  \centering
  \includegraphics[width=5.5in, keepaspectratio=true]{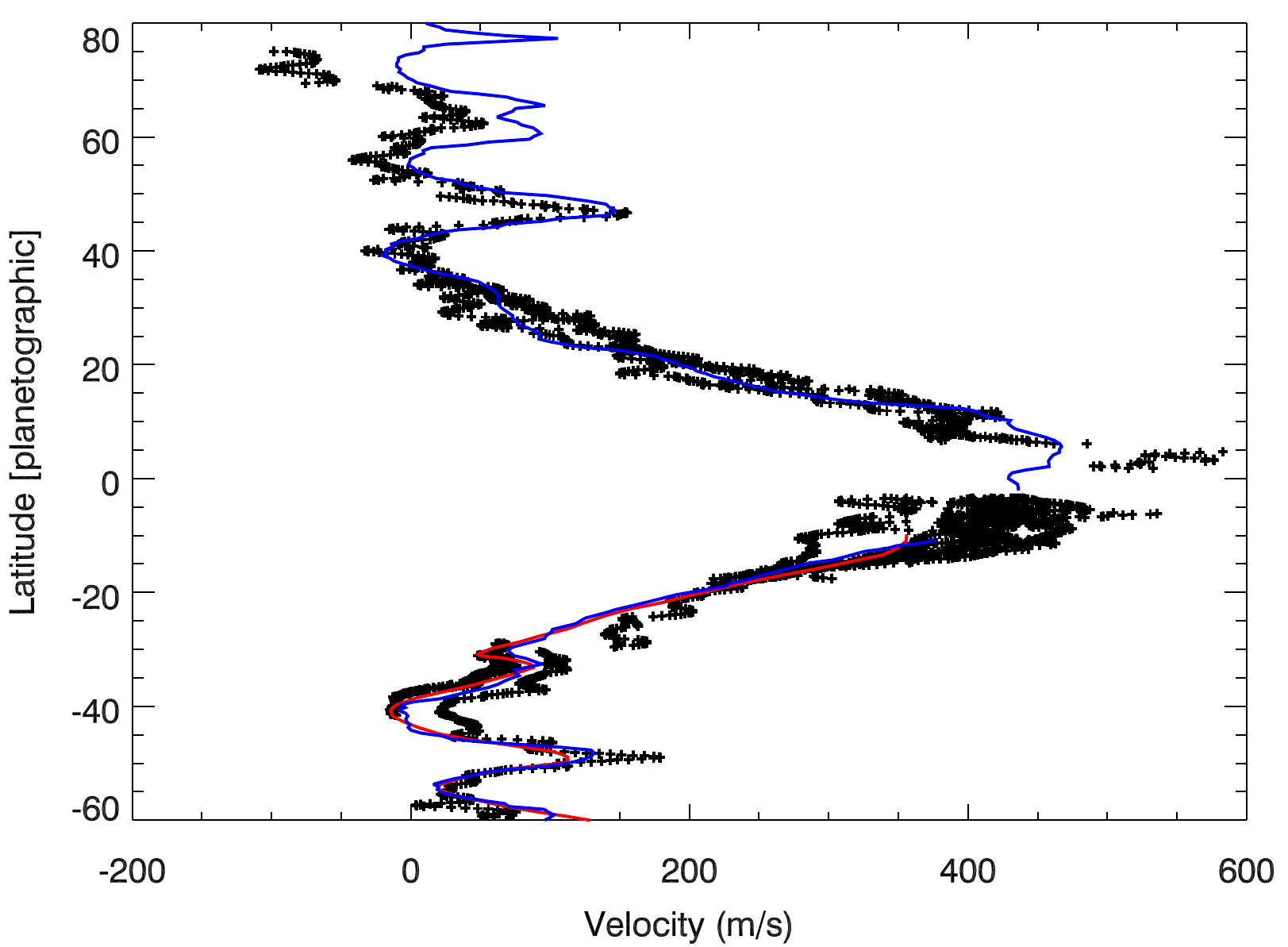}
  \caption[Zonal Wind Profile of Saturn's Atmosphere at Depth]{
    \label{Figure: vims_zprof}
    Zonal wind profile for Saturn's atmosphere constructed from an automated feature tracker analysis of VIMS data. Our VIMS results are shown as small plusses. For comparison, two other profiles are shown: the blue line is from \emph{Voyager} green-filter images \citep{Sanchez-Lavega00}, and the red line is from \emph{Cassini} ISS images\citep{Vasavada06}. Uncertainties for the VIMS measurements are $\sim$10 m s$^{-1}$, and are dependent on the amount of latitudinal wind shear around the measurement.}
\end{figure}

\begin{figure}[htp]
  \centering
  \includegraphics[width=6.5in, keepaspectratio=true]{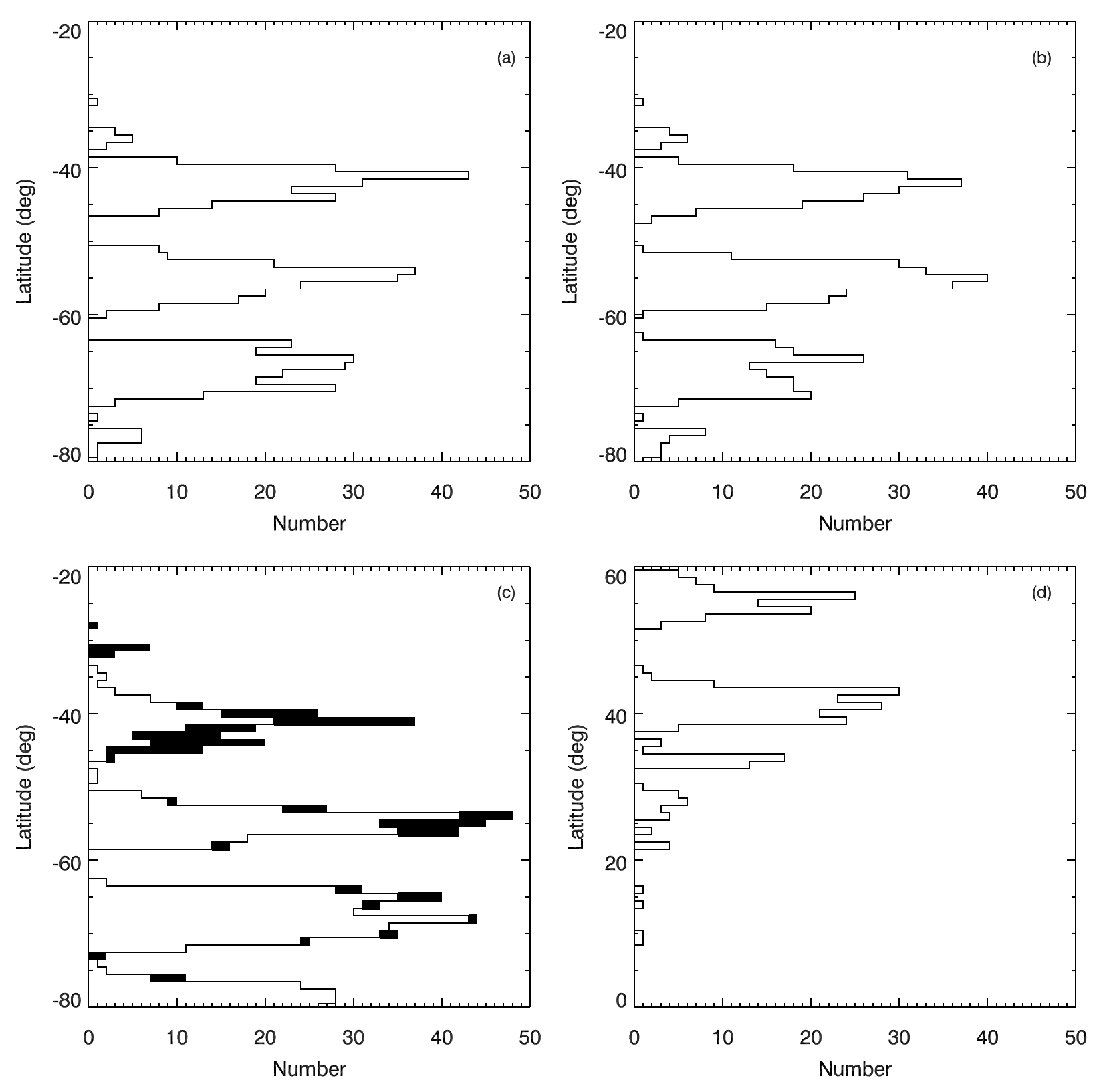}
  \caption[Histogram: Number of Spot Features vs. Latitude]{
    \label{Figure: hist_lat}
    Histogram of the number of spot features as a function of latitude (planetographic). \textbf{a.} VIMS Southern Hemisphere Mosaic I. \textbf{b.} VIMS Southern Hemisphere Mosaic II. \textbf{c.} ISS Southern Hemisphere Mosaic, captured during the first orbit of \emph{Cassini} around Saturn in 2004. The unfilled areas are light spots analyzed in this current study, whereas black bars are dark, low-albedo spot features analyzed by \citet{Vasavada06}. \textbf{d.} VIMS Northern Hemisphere Mosaic.}
\end{figure}

\begin{figure}[htp]
  \centering
  \includegraphics[width=6.5in, keepaspectratio=true]{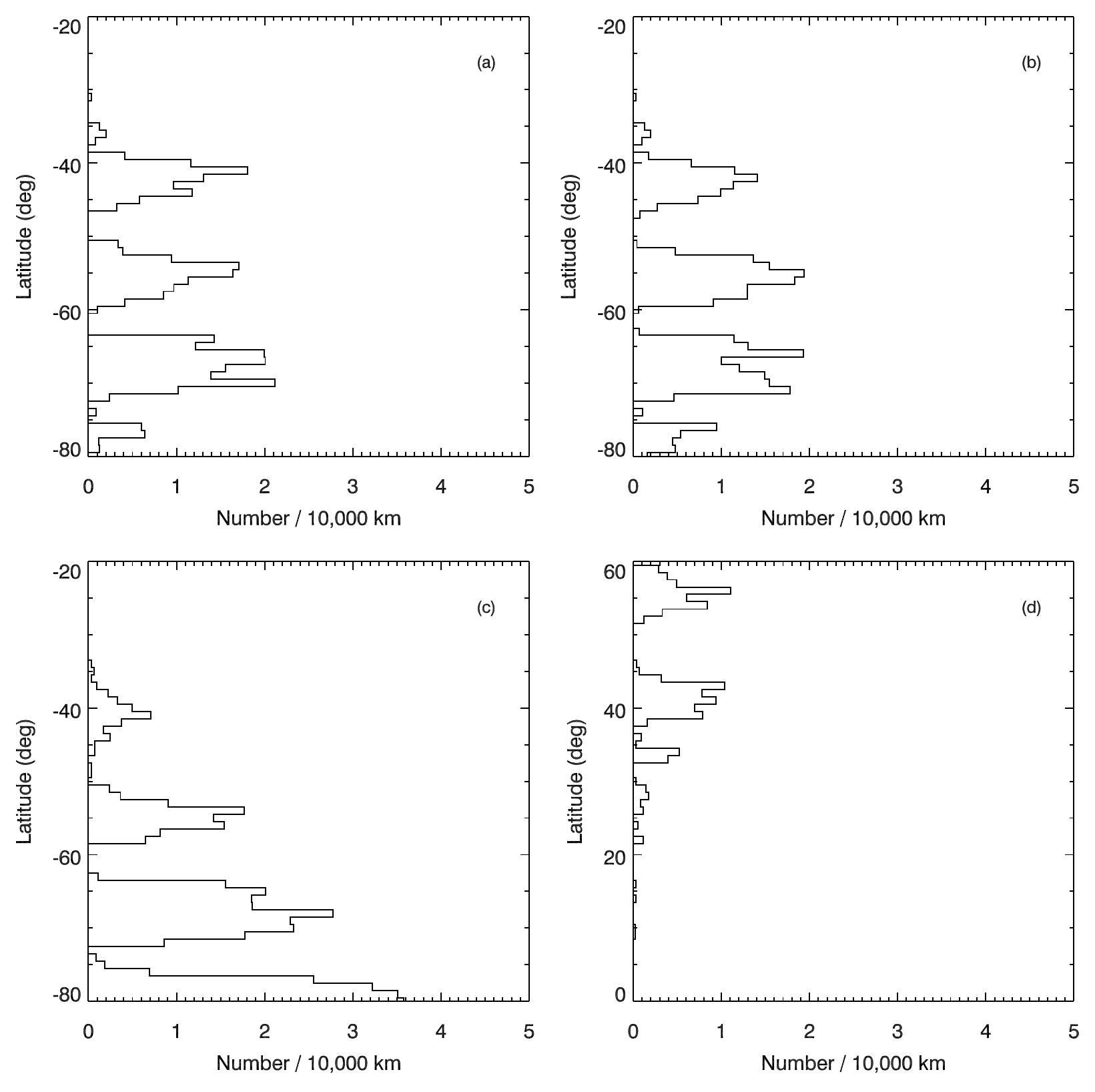}
  \caption[Histogram: Number (Normalized) of Spot Features vs. Latitude]{
    \label{Figure: hist_num_nrml_vs_lat}
    Histogram of the number of spot features per 10,000 km as a function of latitude (planetographic). \textbf{a.} VIMS Southern Hemisphere Mosaic I. \textbf{b.} VIMS Southern Hemisphere Mosaic II. \textbf{c.} ISS Southern Hemisphere Mosaic, captured during the first orbit of \emph{Cassini} around Saturn in 2004. (Both light and dark spots are represented in plot c.) \textbf{d.} VIMS Northern Hemisphere Mosaic.}
\end{figure}

\begin{figure}[htp]
  \centering
  \includegraphics[width=6.5in, keepaspectratio=true]{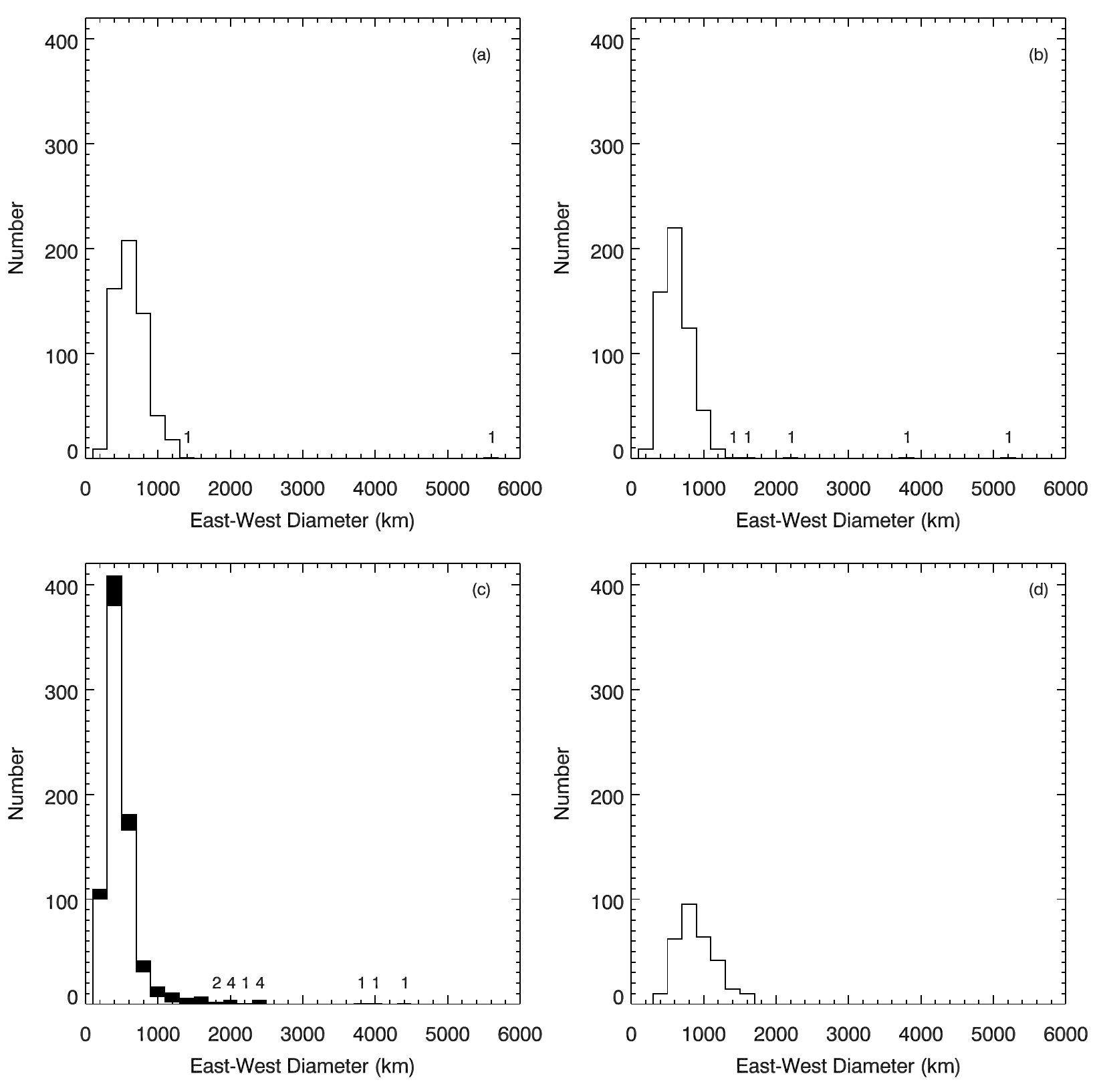}
  \caption[Histogram: Number of Spot Features vs. Horizontal (east-west) span.]{
    \label{Figure: hist_ewdist}
    Histogram of the number of spot features as a function of their horizontal (east-west) span. For clarity, we note the number for small bars that are difficult to distinguish. \textbf{a.} VIMS Southern Hemisphere Mosaic I. \textbf{b.} VIMS Southern Hemisphere Mosaic II. \textbf{c.} ISS Southern Hemisphere Mosaic. The black bars show the contribution from the features analyzed by \citet{Vasavada06}. \textbf{d.} VIMS Northern Hemisphere Mosaic.}
\end{figure}

\begin{figure}[htp]
  \centering
  \includegraphics[width=6.5in, keepaspectratio=true]{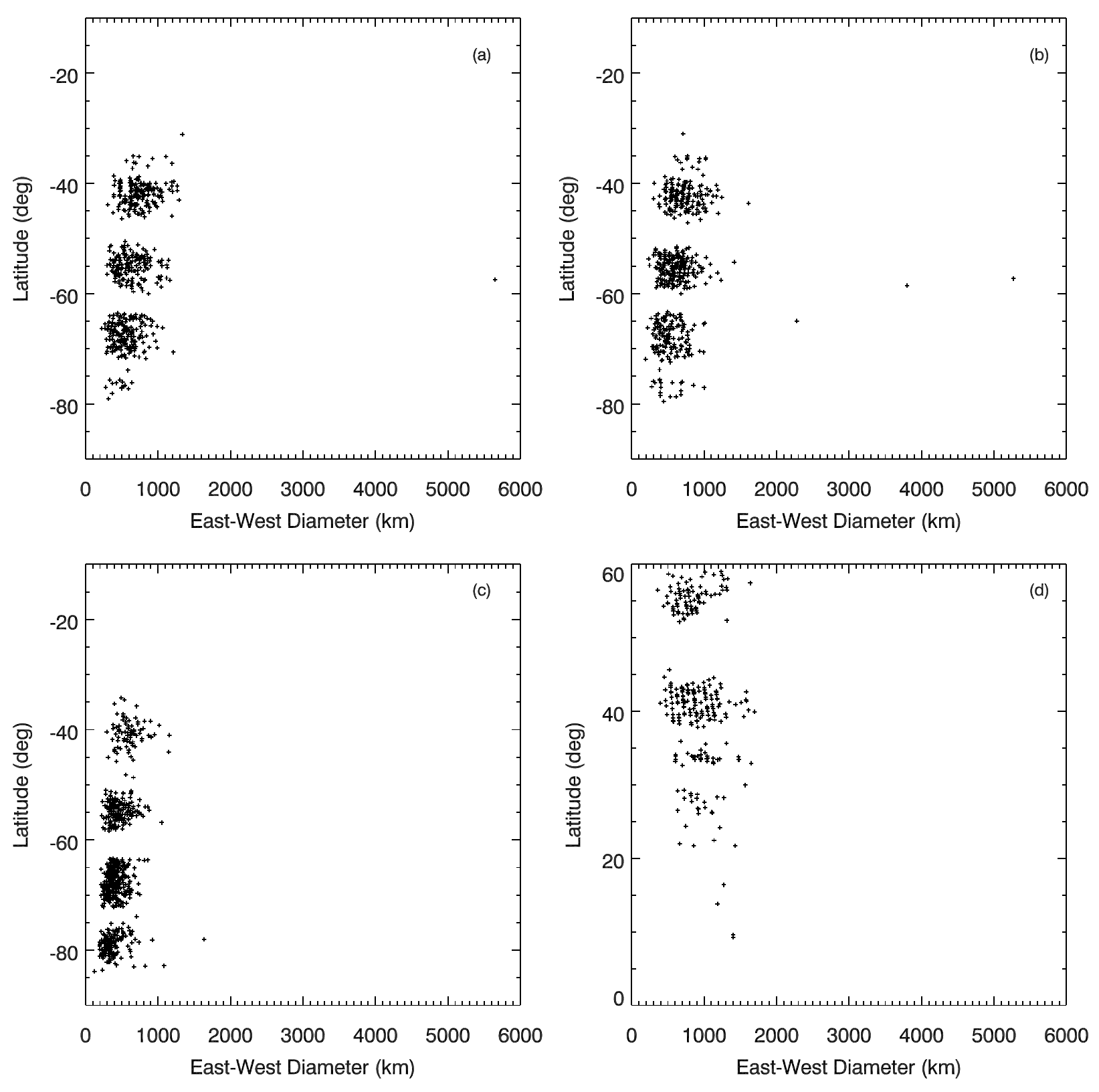}
  \caption[Scatter plot showing a east-west span as a function of planetographic latitude for the spot features]{
    \label{Figure: distvslat}
    Scatter plot showing a spot's east-west span as a function of its latitude (planetographic). \textbf{a.} VIMS Southern Hemisphere Mosaic I. \textbf{b.} VIMS Southern Hemisphere Mosaic II. \textbf{c.} ISS Southern Hemisphere Mosaic. \textbf{d.} VIMS Northern Hemisphere Mosaic.}
\end{figure}


\begin{table}
\begin{center}
\begin{tabular}{|c||c|c||c|c||c|c|}
\hline
& \multicolumn{2}{c||}{\textbf{VIMS S. Hem I}} & \multicolumn{2}{c||}{\textbf{VIMS S. Hem II}} & \multicolumn{2}{c|}{\textbf{ISS S. Hem}} \\
 \hline
Latitude & $N$ & Avg. Diameter (km) & $N$ & Avg. Diameter (km) & $N$ & Avg. Diameter (km)\\
\hline \hline
$<$ 50$^{\circ}$S & 196 & 734.9 $\pm$ 203.5 & 188 & 718.6 $\pm$ 188.3 & 89 & 598.6 $\pm$ 170.4\\ 
50$^{\circ}$S-62$^{\circ}$S & 180 & 634.0 $\pm$ 200.5 & 211 & 609.4 $\pm$ 182.3 & 179 & 462.0 $\pm$ 139.0\\
62$^{\circ}$S-75$^{\circ}$S & 187 & 549.8 $\pm$ 175.2 & 150 & 525.0 $\pm$ 166.0 & 272 & 414.2 $\pm$ 113.9\\
$>$ 75$^{\circ}$S & 14 & 455.3 $\pm$ 109.6 & 19 & 527.0 $\pm$ 198.3 & 146 & 377.8 $\pm$ 145.3\\
\hline
All & 577 & 636.7 $\pm$ 207.9 & 568 & 620.5 $\pm$ 196.0 & 686 & 442.8 $\pm$ 151.2\\
\hline
\end{tabular}

\vspace{15mm}

\begin{tabular}{|c||c|c|}
\hline
 & \multicolumn{2}{c|}{\textbf{VIMS N. Hem}} \\
 \hline
Latitude & $N$ & Avg. Diameter (km) \\
\hline \hline
20$^{\circ}$N-30$^{\circ}$N & 24 & 937.2 $\pm$ 212.5 \\ 
30$^{\circ}$N-37$^{\circ}$N & 33 & 977.5 $\pm$ 242.8 \\
37$^{\circ}$N-46$^{\circ}$N & 136 & 860.5 $\pm$ 223.0 \\
46$^{\circ}$N-60$^{\circ}$N & 90 & 817.0 $\pm$ 223.8 \\
\hline
All & 283 & 866.8 $\pm$ 229.5 \\
\hline
\end{tabular}

\caption{\label{Table: diam_stats} Average east-west diameters for the spot features in the northern and southern hemisphere mosaics, divided by latitude groups. Features with east-west diameters greater than 1500 km have been excluded from this analysis.}
\end{center}
\end{table}

\begin{figure}[htp]
  \centering
  \includegraphics[width=6.5in, keepaspectratio=true]{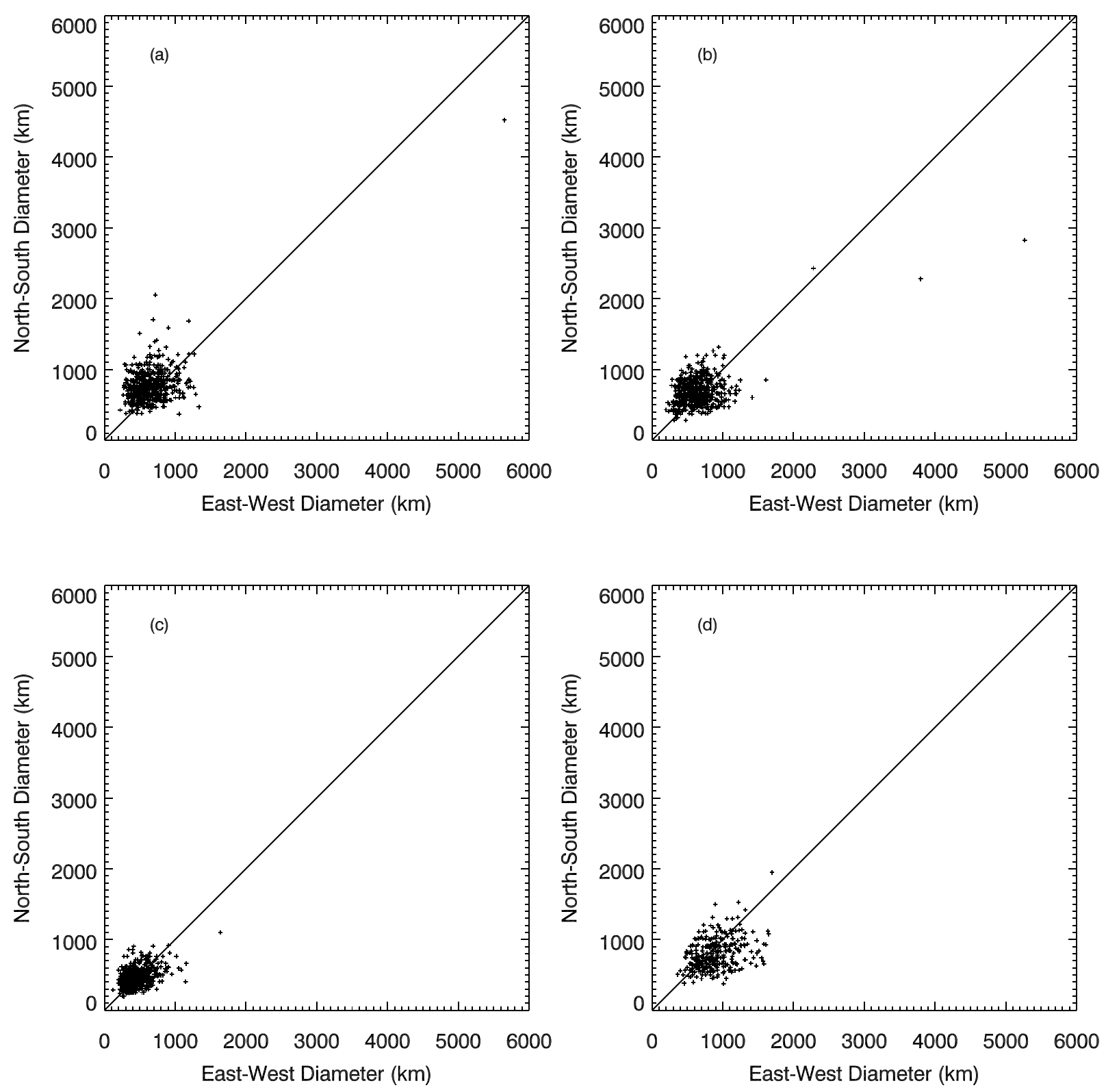}
  \caption[Aspect Ratio (east-west span vs. north-south span) for Spot Features]{
    \label{Figure: aspectratio}
    Plot comparing a spot feature's east-west span versus its north-south span. A line representing an equal aspect ratio is drawn for reference in each plot. \textbf{a.} VIMS Southern Hemisphere Mosaic I. \textbf{b.} VIMS Southern Hemisphere Mosaic II. \textbf{c.} ISS Southern Hemisphere Mosaic. \textbf{d.} VIMS Northern Hemisphere Mosaic.}
\end{figure}

\begin{figure}[htp]
  \centering
  \includegraphics[width=6.5in, keepaspectratio=true]{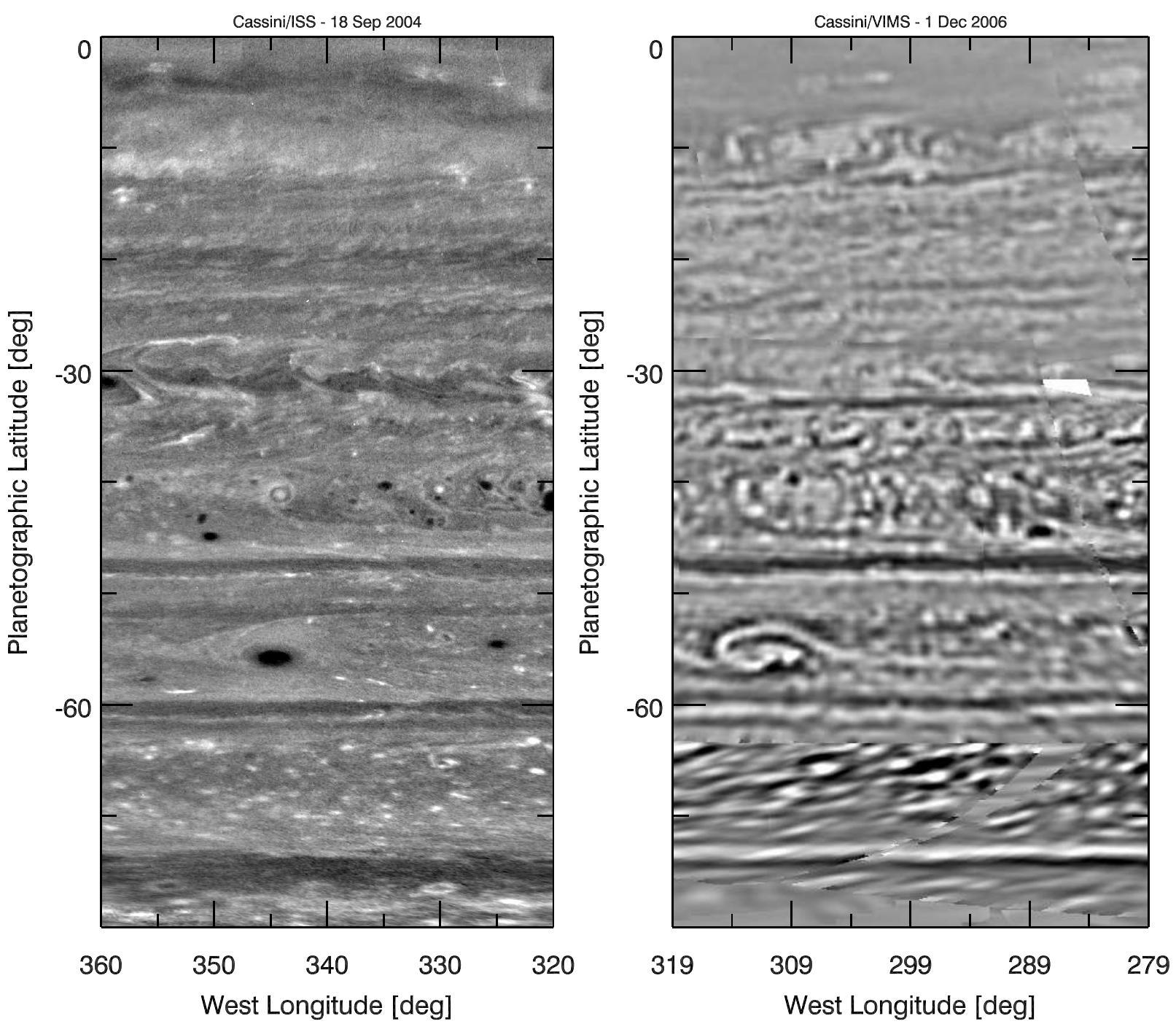}
  \caption[Comparison of ISS and VIMS mosaics]{
    \label{Figure: vims_and_iss}
    \textbf{(left)} Portion of the southern hemisphere ISS mosaic published as Figure 1 in \citet{Vasavada06}. The contrast has been increased to enhance the observed features. \textbf{(right)} \emph{Inverted} portion of the southern hemisphere VIMS mosaic seen in Figure \ref{Figure: mosaic_southhem1}.}
\end{figure}

\end{document}